\documentclass[journal]{IEEEtran}
\usepackage{graphicx}  % Required for inserting images
\usepackage{amsmath}
\usepackage{amssymb}
\usepackage{hyperref}
\usepackage{listings}
\usepackage{xcolor}
\usepackage{url}

\newcommand{\nclus}{\ensuremath{n_\text{clus}}}
\newcommand{\ncomp}{\ensuremath{n_\text{comp}}}
\newcommand{\nanode}{\ensuremath{n_\text{anode}}}
\newcommand{\nedge}{\ensuremath{n_\text{edge}}}

\definecolor{mygreen}{rgb}{0,0.6,0}
\definecolor{mygray}{rgb}{0.5,0.5,0.5}

\title{Data-driven optimization of pixelated CdZnTe spectrometers for uranium enrichment assay}
\author{
    Jayson~R.~Vavrek, Thomas~D.~MacDonald, Hannah~S.~Parrilla, Nikhil~S.~Deshmukh, Mital~A.~Zalavadia, Benjamin~S.~McDonald
    \thanks{
        J.R.~Vavrek, T.D.~MacDonald, and H.S.~Parrilla are with the Nuclear Science Division, Lawrence Berkeley National Laboratory, Berkeley, CA, 94720, USA.
        N.S.~Deshmukh, M.A.~Zalavadia, and B.S.~McDonald are with the National Security Directorate, Pacific Northwest National Laboratory, Richland, WA, 99354, USA.

        The work presented in this paper was funded by the National Nuclear Security Administration of the Department of Energy, Office of International Nuclear Safeguards.
        This work was performed under the auspices of the U.S.\ Department of Energy by Lawrence Berkeley National Laboratory (LBNL) under Contract DE-AC02-05CH11231.
    }
}
\date{\today}

\begin{document}

\maketitle

\begin{abstract}
In recent work [Vavrek et al.~(2025)], we developed the performance optimization framework {\tt spectre-ml} for gamma spectrometers with variable performance across many readout channels.
The framework uses non-negative matrix factorization (NMF) and clustering to learn groups of similarly-performing channels and sweep through various learned channel combinations to optimize the performance tradeoff of including worse-performing channels for better total efficiency.
In this work, we integrate the {\tt pyGEM} uranium enrichment assay code with our {\tt spectre-ml} framework, and show that the U-235 enrichment relative uncertainty can be directly used as an optimization target.
We find that this optimization reduces relative uncertainties after a $30$-minute measurement by an average of $20\%$, as tested on six different H3D M400 CdZnTe spectrometers, which can significantly improve uranium non-destructive assay measurement times in nuclear safeguards contexts.
Additionally, this work demonstrates that the {\tt spectre-ml} optimization framework can accommodate arbitrary end-user spectroscopic analysis code and performance metrics, enabling future optimizations for complex Pu spectra.

\end{abstract}

\section{Introduction}
\begin{figure}[!htbp]
    \centering
    \includegraphics[width=1.00\linewidth]{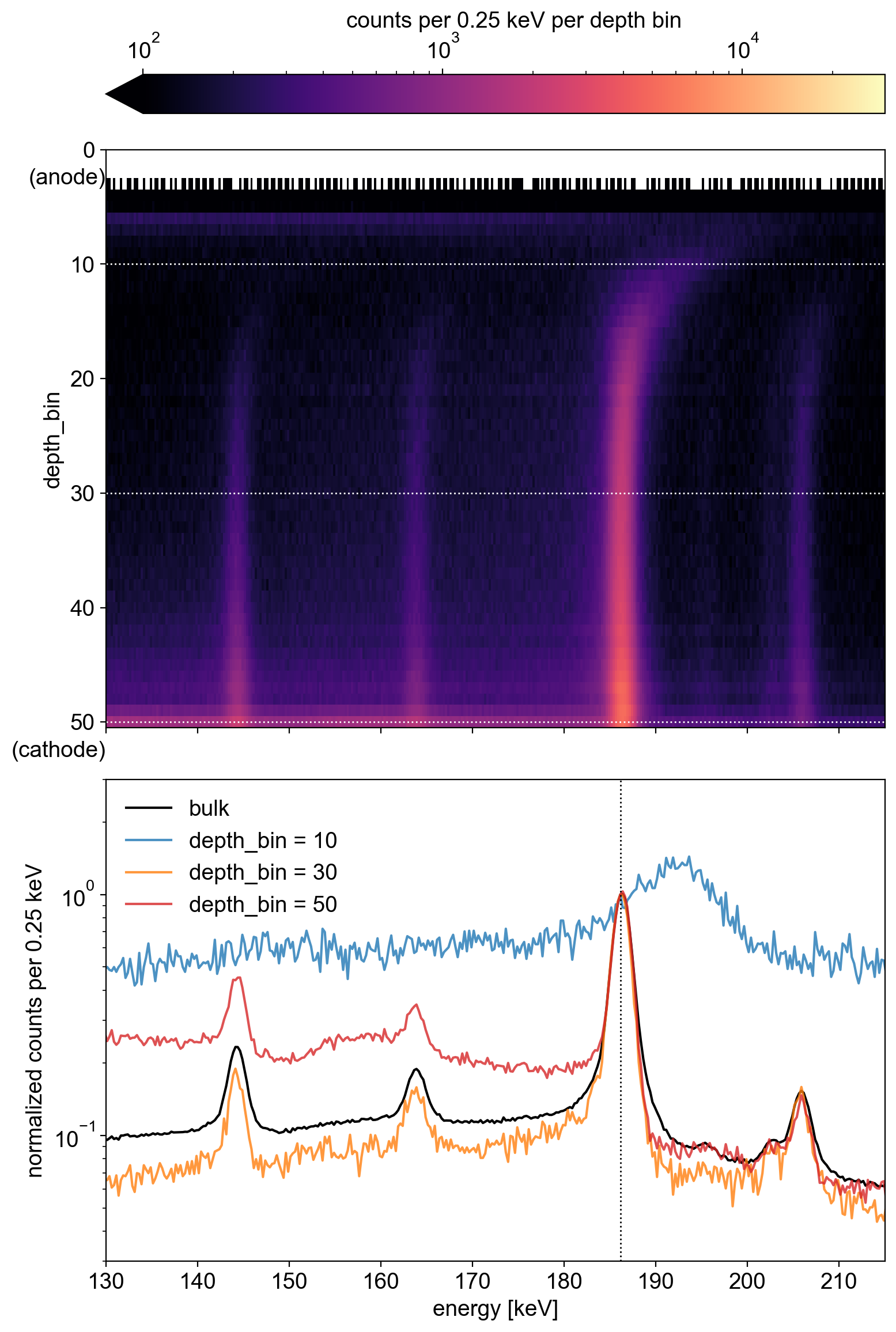}
    \caption{
        Uranium gamma spectra as a function of depth in the PNNL M400 detector.
        Top: heatmap of all depth spectra.
        The white dotted lines indicate the depth bin selections used in the bottom panel.
        Bottom: bulk spectrum and spectra at three depth bin selections, each normalized to their intensity at $186$~keV (black dotted line) to better show their shapes.
        Depth bins near the anode degrade in resolution and can exhibit gain drifts, while depth bins near the cathode have increased backgrounds.
    }
    \label{fig:depth_spectra}
\end{figure}

The H3D M400 gamma spectrometer~\cite{m400_spec_sheet} is being adopted by the International Atomic Energy Agency (IAEA) as its primary in-field uranium enrichment non-destructive assay (NDA) technology~\cite{iaea2022disp, lebrun2022next, dodane2023large}.
The M400 features four CdZnTe (CZT) crystals each pixelated to an $11 \times 11$ grid, and offers medium resolution gamma spectroscopy at room temperature in a compact form factor.
In previous work, we showed that the spatial variations in detector performance~\cite{goodman2022energy} within the CZT crystal volumes could be exploited to optimize spectroscopic performance metrics~\cite{vavrek2025data}.
In particular, a balance can be found between identifying and rejecting poorly-performing detector regions (e.g., the poor-resolution anode regions and/or poor-signal-to-background cathode regions as shown in Fig.~\ref{fig:depth_spectra}, or pixels with high defect concentrations~\cite{bolotnikov2007cumulative}) and the associated loss of efficiency.
Because of the large number of possible voxel combinations (${\sim}10^{7285}$), computing the globally optimum set of voxels to include is infeasible, and approximate, data-driven methods were developed.
Several of the example optimizations therein focused on minimizing the relative uncertainty in a peak fit amplitude parameter as proofs of concept in lieu of more advanced spectral performance metrics.

In this work, we have integrated our spectroscopic optimization software {\tt spectre-ml} with {\tt GEM}~\cite{berlizov2022gem}, the actual software used by the IAEA to conduct uranium NDA measurements.
We show that the relative uncertainty on the {\tt GEM} enrichment calculation can directly be used as a {\tt spectre-ml} optimization target, i.e., that {\tt spectre-ml} can improve the statistical confidence of the IAEA's in-field enrichment measurements without increasing measurement times, or conversely can achieve the same statistical confidence in shorter measurement times.
Similarly, {\tt spectre-ml} improvements could widen the pool of CZT crystals meeting the performance requirements necessary for in-field measurements, potentially driving down manufacturing costs.
Such software-based optimization could reduce the need for and/or complement various hardware-based advances such as design stage optimizations~\cite[\S 3.2]{li2024research}, \cite{huang2025machine} that do not account for individual crystal differences, digital pulse processing for improved energy resolution~\cite{petryk2025improved}~\cite[\S 5.2]{smith2024summary}, and the use of the more spatially-homogeneous CdZnTeSe~\cite{roy2019evaluation} instead of CZT.

This paper is structured as follows: Section~\ref{sec:methods} covers the integration of the {\tt spectre-ml} and {\tt GEM} codes at a high level, discussing recent improvements to the former to better enable arbitrary end-user workflows, and providing an overview of the {\tt GEM} workflow itself.
It also describes the uranium measurements and optimization parameter sweeps used for analysis.
Section~\ref{sec:results} gives the results of six {\tt GEM}-based {\tt spectre-ml} optimizations for six M400 units, showing that the enrichment relative uncertainty can be directly used as an optimization target.
Section~\ref{sec:discussion} then provides additional interpretation of the results, discusses some limitations, and suggests avenues for further work.
Finally, the Appendix gives a pseudocode overview of how an end-user can integrate their own spectrum analyses into the {\tt spectre-ml} optimization framework.

\section{Methods}\label{sec:methods}

\subsection{Integration of the {\tt spectre-ml} and {\tt pyGEM} codes}
The {\tt spectre-ml}~\cite{vavrek2025data, vavrek2023spectral} code is a Python package for optimizing the performance of many-channel gamma spectrometers by intelligently rejecting data from poorly-performing detector channels (i.e., spatial regions).
It uses non-negative matrix factorization (NMF)~\cite{lee1999learning, wang2012nonnegative} and unsupervised clustering algorithms to learn groups of voxels with similar performance, compute performance metrics across various voxel cluster combinations, and then sweep over hyperparameters such as clustering algorithm, number of clusters, number of NMF components, etc., to find the best overall set of voxels to include.
Previous versions of {\tt spectre-ml} were restricted to single Doniach~\cite{doniach1970many} or Gaussian peak fit workflows using the {\tt becquerel} library for peak fits.
In this work we have generalized the {\tt spectre-ml} software to abstract out the dependence on {\tt becquerel} and enable arbitrary user-defined spectrum analyses, in particular those based on {\tt GEM}.
The {\tt pyGEM}~\cite{mcdonald2022gamma} code is a Python interface to the General Enrichment Measurements ({\tt GEM}) code~\cite{berlizov2022gem} used by the IAEA for uranium enrichment NDA.
It fits measured spectra in the ${\sim}120$--$270$~keV region of interest using a set of known U-235 emission lines, forward scatter, and high-energy downscatter profiles, and has recently been updated to handle the asymmetric peak shapes of CZT.
In contrast to our Doniach fit {\tt becquerel} workflow, {\tt pyGEM} uses a triple-Gaussian model for each peak in CZT---one primary peak, one low-energy tail, and one high-energy tail.
The user provides {\tt pyGEM} a spectrum from a sample of known U-235 enrichment, and it will then compute a linear calibration between the fit net count rate in the $185.7$~keV peak and the U-235 atom abundance that can be used to compute the enrichment from an unknown sample spectrum.
It is also possible to incorporate correction factors for certain changes between the calibration and sample spectra such as changes in the sample matrix (e.g., U metal vs.\ U$_3$O$_8$), Al and steel wall thicknesses, etc.
The {\tt pyGEM} enrichment relative uncertainty calculation incorporates both statistical and systematic (i.e., fit error) components, and thus the {\tt spectre-ml} optimization can be viewed as trading off statistical vs.\ systematic uncertainties as voxel clusters with higher fit error are removed.

\begin{figure*}[!htbp]
    \centering
    \includegraphics[width=1.0\linewidth]{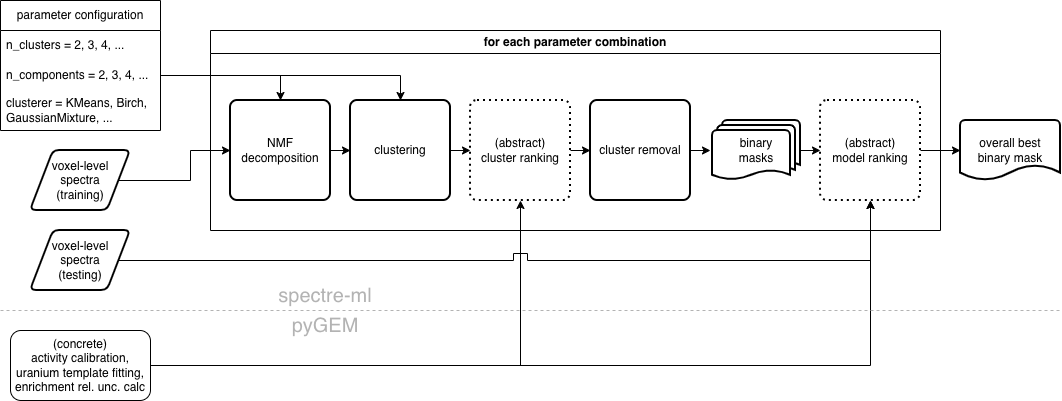}
    \caption{
        Overview of the {\tt spectre-ml} + {\tt pyGEM} pipeline, adapted from Fig.~2 of Ref.~\cite{vavrek2025data}.
    }
    \label{fig:pipeline}
\end{figure*}

To integrate the two codes, some {\tt spectre-ml} updates were required to handle both the {\tt becquerel} and {\tt pyGEM} workflows.
As shown in the pseudocode in the Appendix, {\tt spectre-ml} now provides the abstract base classes {\tt SpectrumAnalyzer} and {\tt RankingMetric}.
The user must provide concrete subclasses implementing {\tt SpectrumAnalyzer.analyze} and {\tt RankingMetric.calc}, where the latter method takes a spectrum analyzed (e.g., fit) by the former and extracts a single {\tt float} metric such as the relative uncertainty in a fit parameter ({\tt becquerel}) or in an enrichment ({\tt pyGEM}) calculation.
The usual {\tt spectre-ml} analysis then proceeds and the various voxel selections tested are then ranked by these metric values, with the best metric value indicating which voxel selection to use for the given application defined by the input spectra and choice of spectrum analysis and ranking metric.
An overview of the pipeline is given in Fig.~\ref{fig:pipeline}.
We note that no {\tt pyGEM} code changes were required, indicating that {\tt spectre-ml} end-users can relatively easily slot in their own analysis routines.

\subsection{Uranium enrichment assay optimization}
Multiple U$_3$O$_8$ standards measurements were performed with six US National Laboratory M400 detectors (Brookhaven, Idaho, Los Alamos, Oak Ridge, Pacific Northwest, and Sandia National Laboratories)~\cite{smith2024summary} and provided to Lawrence Berkeley National Laboratory for analysis with {\tt spectre-ml}.
In this work we use the $30$-minute collimated M400 measurements of Standard Reference Material (SRM) 969~\cite{srm969} samples with label U-235 enrichments of $1.94$~wt\% and $4.46$~wt\% (certified values of $(1.9420 \pm 0.0014)$~wt\% and $(4.4623 \pm 0.0032)$~wt\%).
The latter is used as the known-enrichment sample for the {\tt pyGEM} activity calibration while the former is used as the unknown sample to be assayed.

The relative uncertainty in the U-235 enrichment is a crucial performance metric for U-235 NDA as it is directly used in computing the \textit{operator-inspector difference} (OID) in terms of $z$-scores or ``sigmas'', and used to compare uncertainties against the International Target Values (ITVs) for the GEM-based verification method~\cite{iaea2022itvs}.
A lower relative uncertainty on the measured U-235 enrichment increases the statistical power of the measurement to detect diversions from declared enrichments and thus potential treaty violations.
Since the statistical component of the relative uncertainty decreases with the square root of measurement time, long measurement times may be required for high statistics; longer measurement times are however not always an option for inspectors performing verifications in the field.
Optimizing for lower relative uncertainty can therefore improve the time-efficiency of IAEA NDA tasks.

It is also useful to quantify improvements in the {\tt pyGEM} fit quality after voxel clusters are removed.
A {\tt pyGEM} analysis typically uses the ``{\tt Nqfit}'' metric, an ``intensity-normalized'' version of the standard reduced chi-squared $\chi_\nu^2$, given by~\cite{iaea2020physical}
\begin{align}
    {\tt Nqfit} - 1 = \left( \chi_\nu^2 - 1 \right) \frac{100 \times 10^3}{{\tt sum} + 50 \times 10^3}.
\end{align}
Here {\tt sum} is the total counts of the U-235 fit component, such that {\tt Nqfit} is essentially a $\chi_\nu^2$ scaled to an assumed ``normal'' counting statistics of $50 \times 10^3$ counts.

In the optimization examples of Section~\ref{sec:results}, the {\tt spectre-ml} optimization considers four different classes of algorithms for creating voxel clusters (see also Ref.~\cite{vavrek2025data}):
\begin{enumerate}
    \item machine learning (ML): standard machine learning clustering algorithms, e.g., Gaussian Mixture, Agglomerative, or $k$-means clustering, using the NMF spectral decomposition weights as inputs;
    \item heuristic:
    \begin{enumerate}
        \item equal-depth: segments the detector into $\nclus$ equally-sized regions in depth;
        \item edge-and-anode: segments the detector into three regions: a depth region of $\nanode$ depth bins, an edge region of the $\nedge$ outer pixels of each of the four detector modules (not including the anode depth bins), and the rest of the detector volume;
    \end{enumerate}
    \item greedy: computes the metric directly on spectra within every distinct spatial element (voxel, pixel, depth bin, or crystal), ranks each element by its individual metric, then successively clusters elements together in that order;
    \item random: assigns $\nclus$ random cluster labels at the voxel, pixel, depth bin, or crystal level.
\end{enumerate}
The ML parameter sweeps consist of $\nclus = 2$--$6$ clusters (Gaussian Mixture only), $\ncomp = 2$--$6$ NMF components, and $\alpha_W = 0.0$ (i.e., no NMF regularization).
The equal-depth and edge-and-anode algorithms use $\nclus = 2$--$6$ and a fixed $(\nedge, \nanode) = (1, 15)$, respectively.
The greedy detector algorithm is used, but not the greedy pixel, depth bin, or voxel algorithms due to memory constraints, and a single sample is run for each of the random pixel, depth bin, and voxel clusterers with $\nclus = 2$--$6$.

\section{Results}\label{sec:results}

\begin{figure*}[!htbp]
    \centering
    \includegraphics[width=0.49\linewidth]{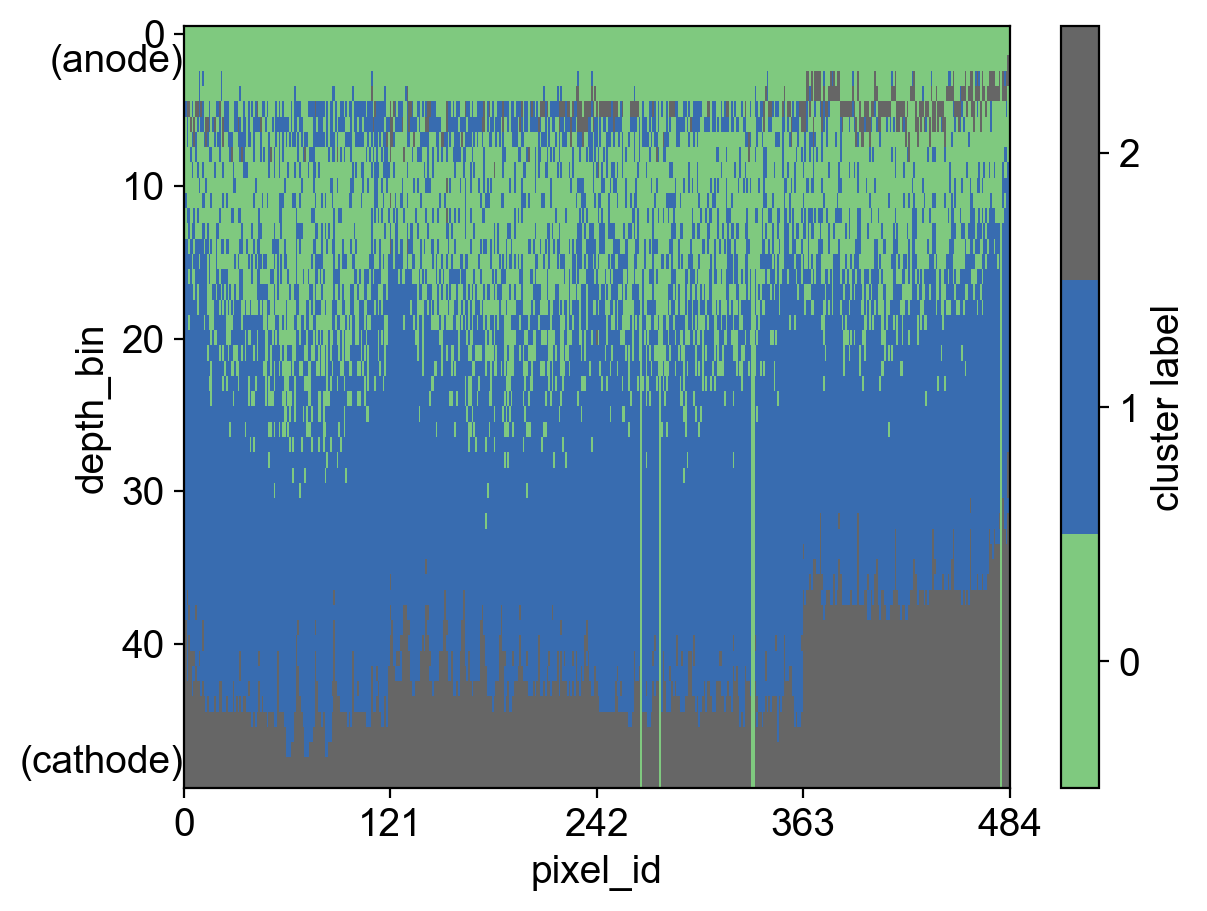}
    \includegraphics[width=0.49\linewidth]{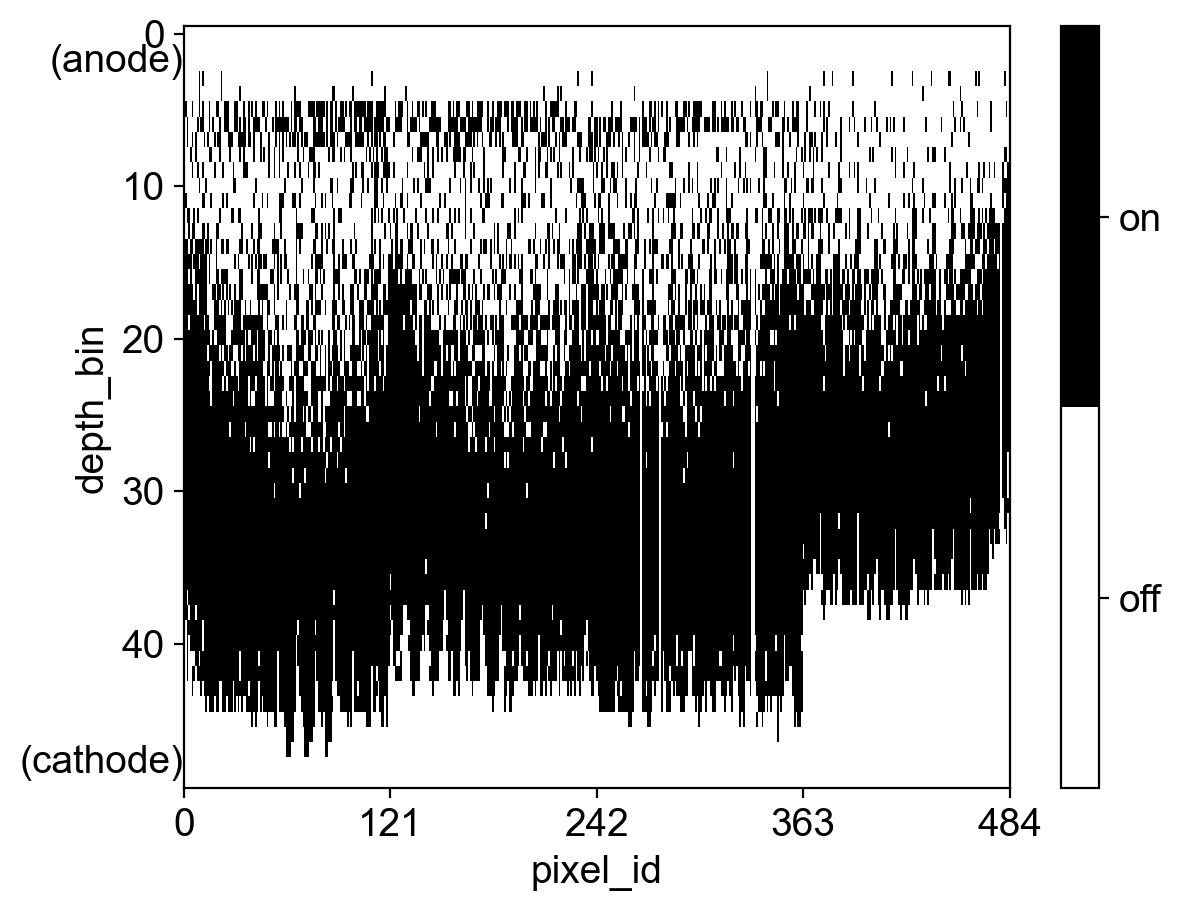}\\
    \includegraphics[width=0.49\linewidth]{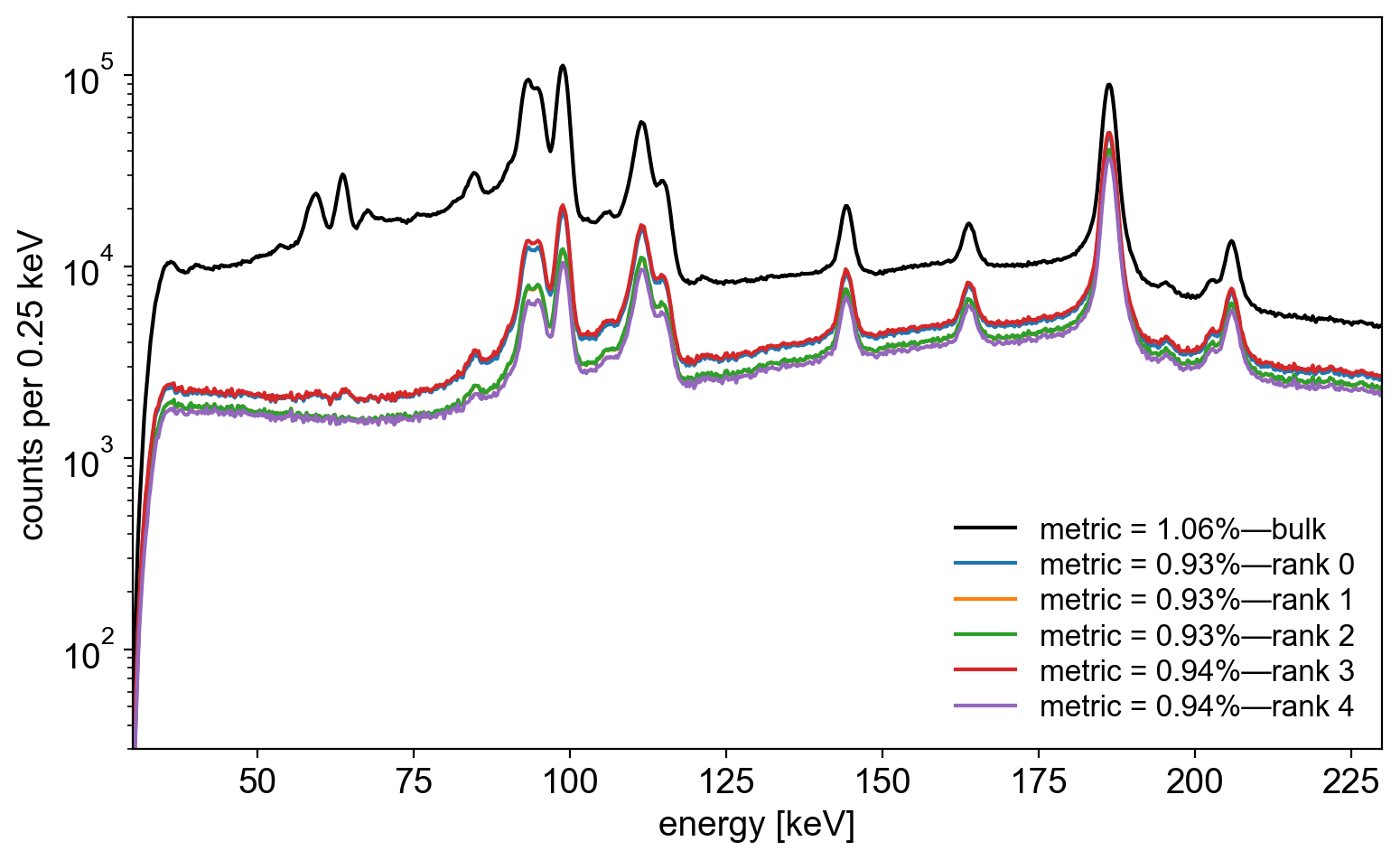}
    \includegraphics[width=0.49\linewidth]{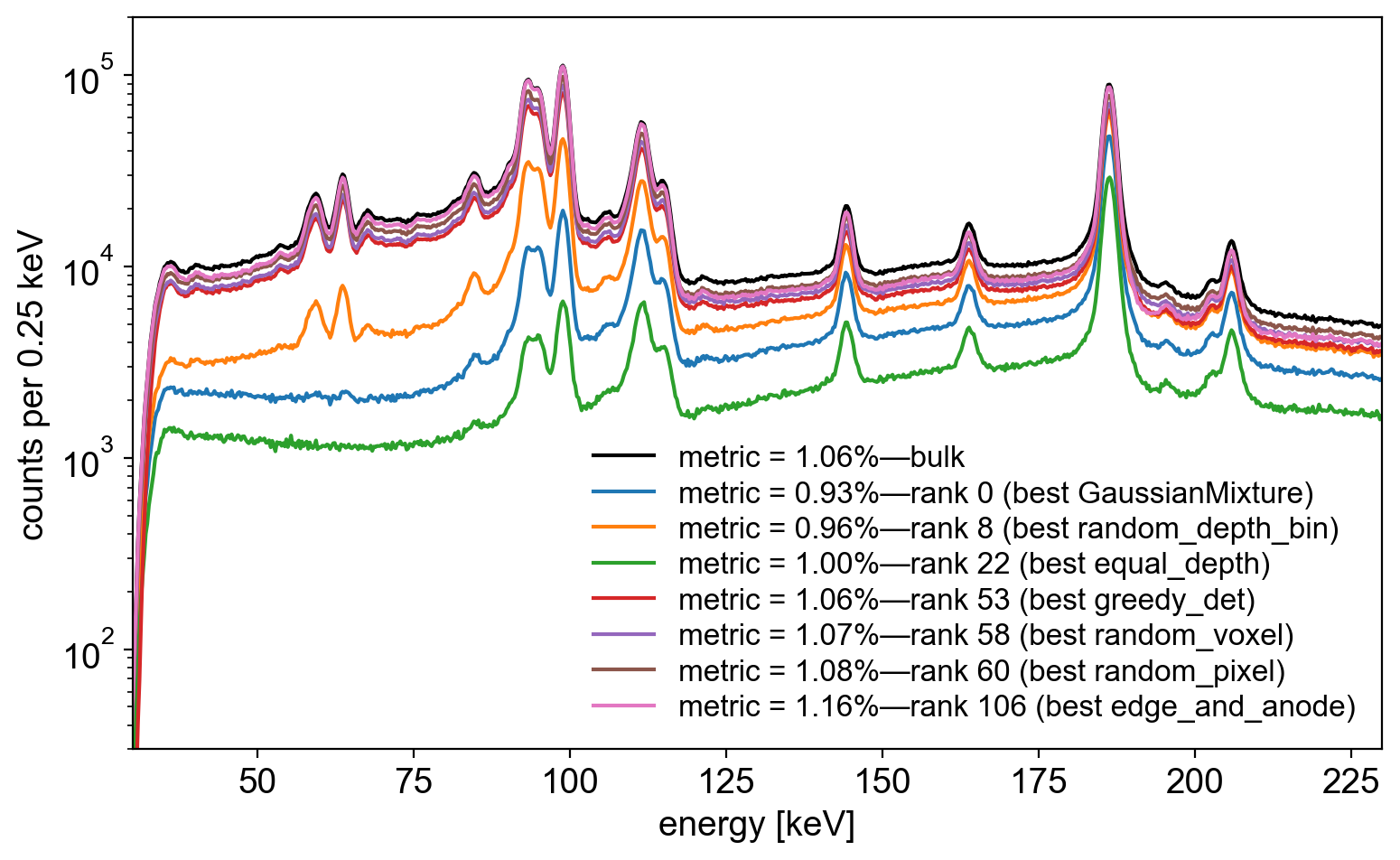}\\
    \includegraphics[width=0.49\linewidth]{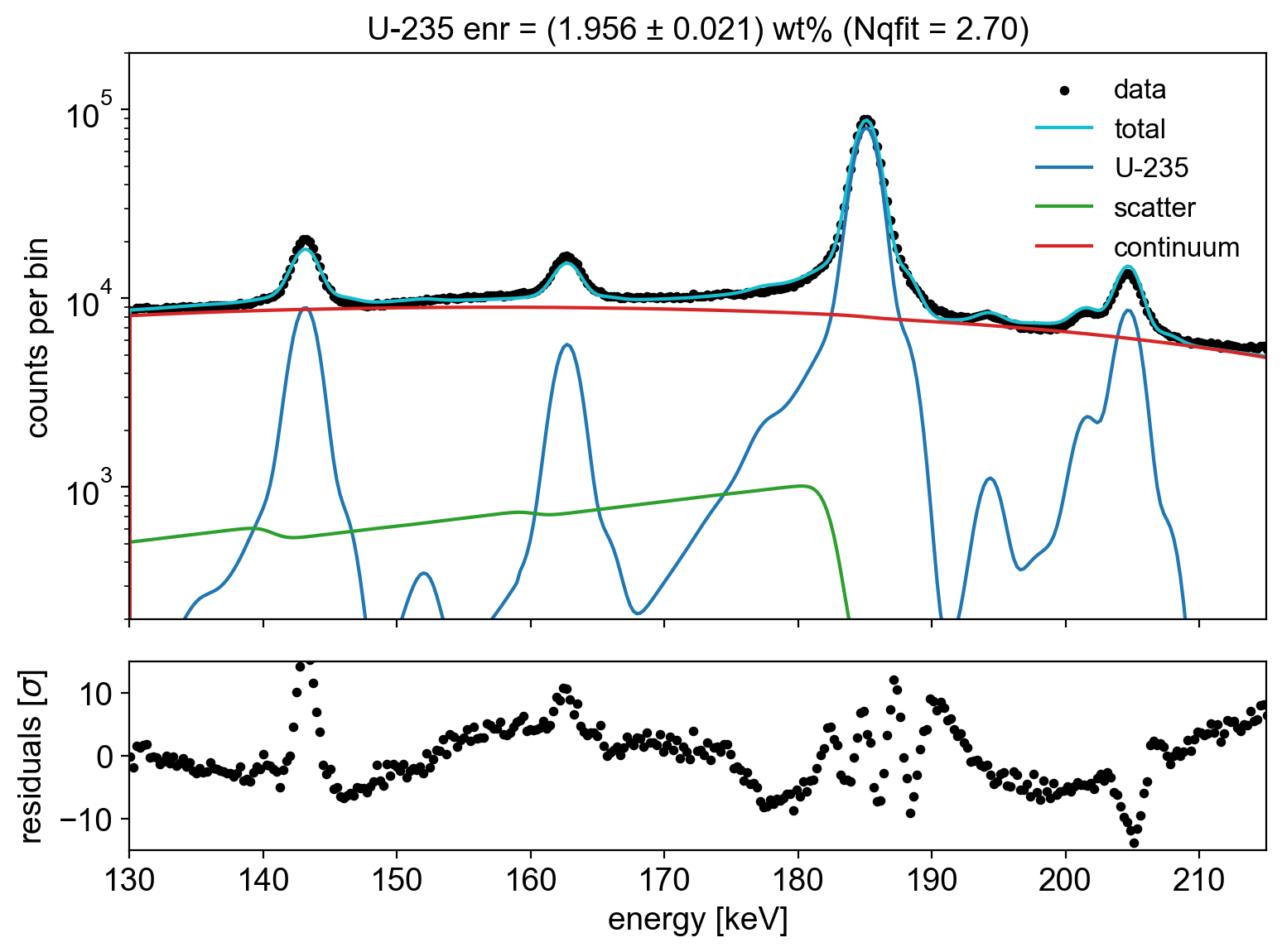}
    \includegraphics[width=0.49\linewidth]{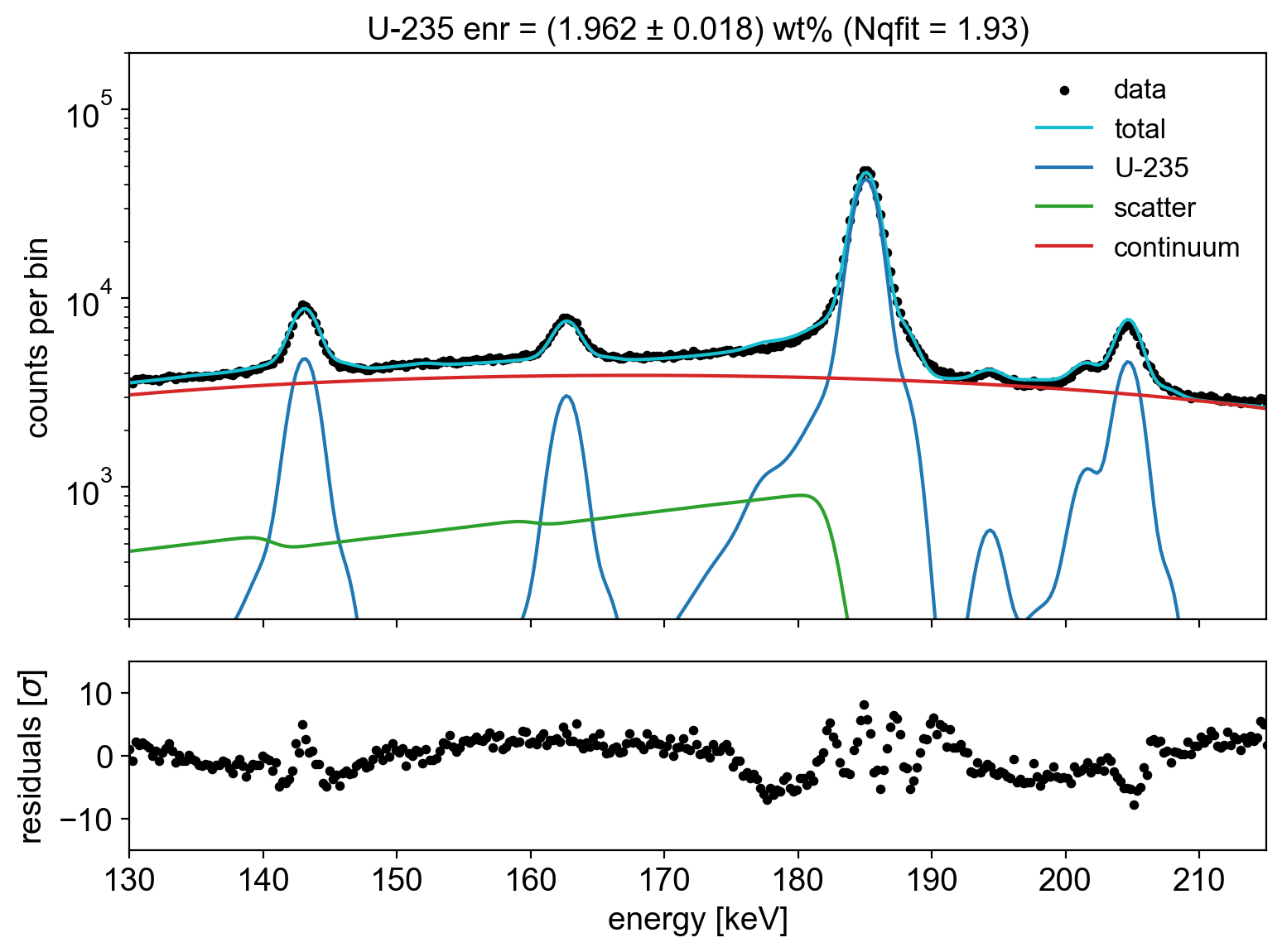}
    \caption{
        Uranium enrichment relative uncertainty optimization example with the PNNL M400 detector.
        Top left: best cluster labels.
        Top right: best cluster mask.
        Center left: top $5$ spectra ranked by enrichment relative uncertainty.
        Center right: best spectrum from each class of clustering algorithm.
        Bottom left: {\tt pyGEM} fit to the bulk spectrum.
        Bottom right: {\tt pyGEM} fit to the best spectrum.
    }
    \label{fig:enrichment}
\end{figure*}

Fig.~\ref{fig:enrichment} shows the results of the {\tt spectre-ml} + {\tt pyGEM} U-235 enrichment relative uncertainty optimization using the PNNL M400 uranium standards dataset.
The enrichment relative uncertainty of $1.06\%$ from the bulk (unoptimized) spectrum is improved to $0.93\%$ (a relative improvement of $12\%$) when using $\ncomp = 3$ and retaining $1$ out of $3$ Gaussian Mixture clusters.
This voxel mask reduces the $185.7$~keV net peak area from $8.22 \times 10^{5}$~cps to $4.45 \times 10^{5}$~cps (an efficiency reduction to $54\%$ of the unoptimized detector) and slightly degrades the peak full width at half maximum (FWHM) from $1.18\%$ to $1.20\%$, but improves the overall fit quality {\tt Nqfit} from $2.70$ to $1.93$ by reducing fit residuals in both continuum and photopeak regions.
The top $5$ masks achieve similar final metrics of $0.93\%$--$0.94\%$.
The Gaussian Mixture clusterer generally outperforms the heuristic clustering algorithms, since it can learn and better represent trends in a particular dataset.
The edge-and-anode and random pixel and voxel clusterers achieve results worse than the bulk value of $1.06\%$, due in part to their inability to learn from the data, but also their more limited parameter sweeps.

\begin{figure*}[!htbp]
    \centering
    \includegraphics[width=0.32\linewidth]{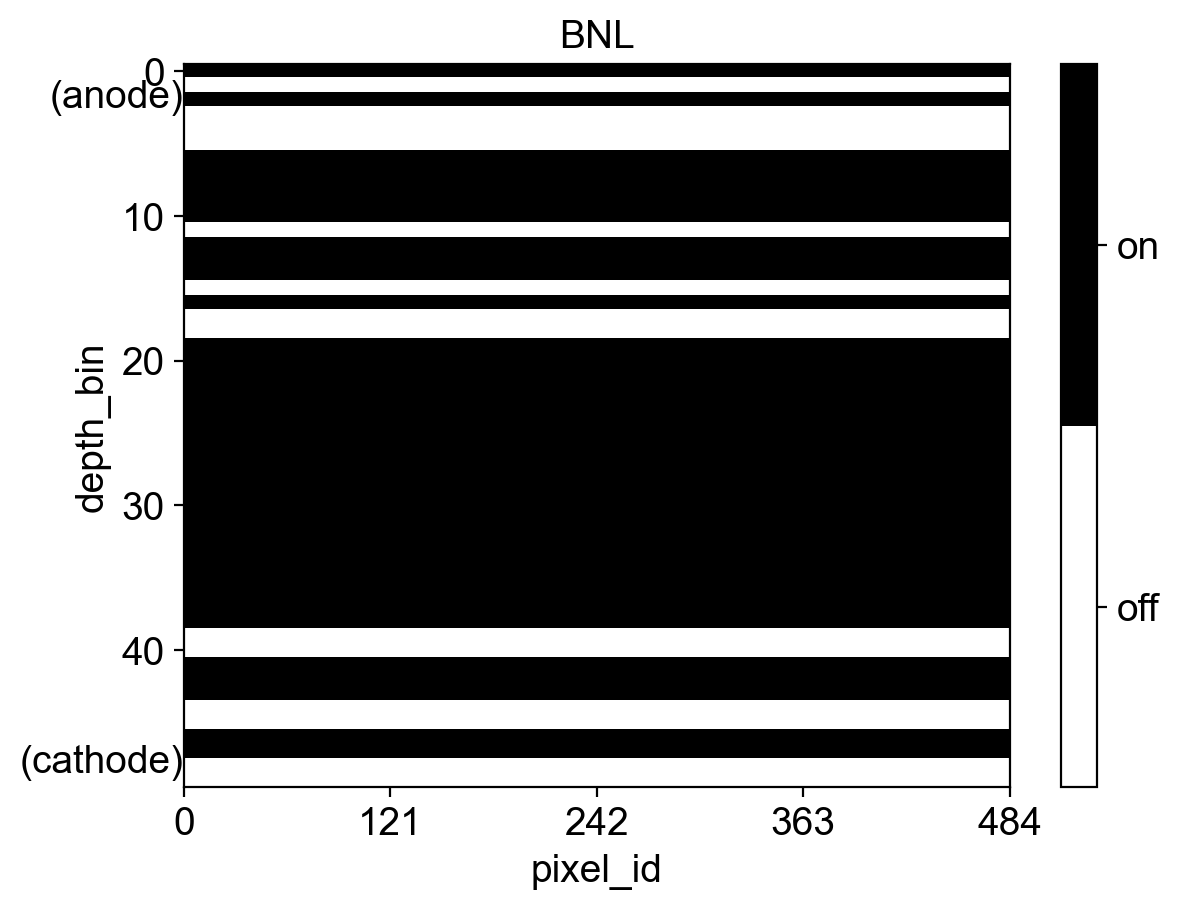}
    \includegraphics[width=0.32\linewidth]{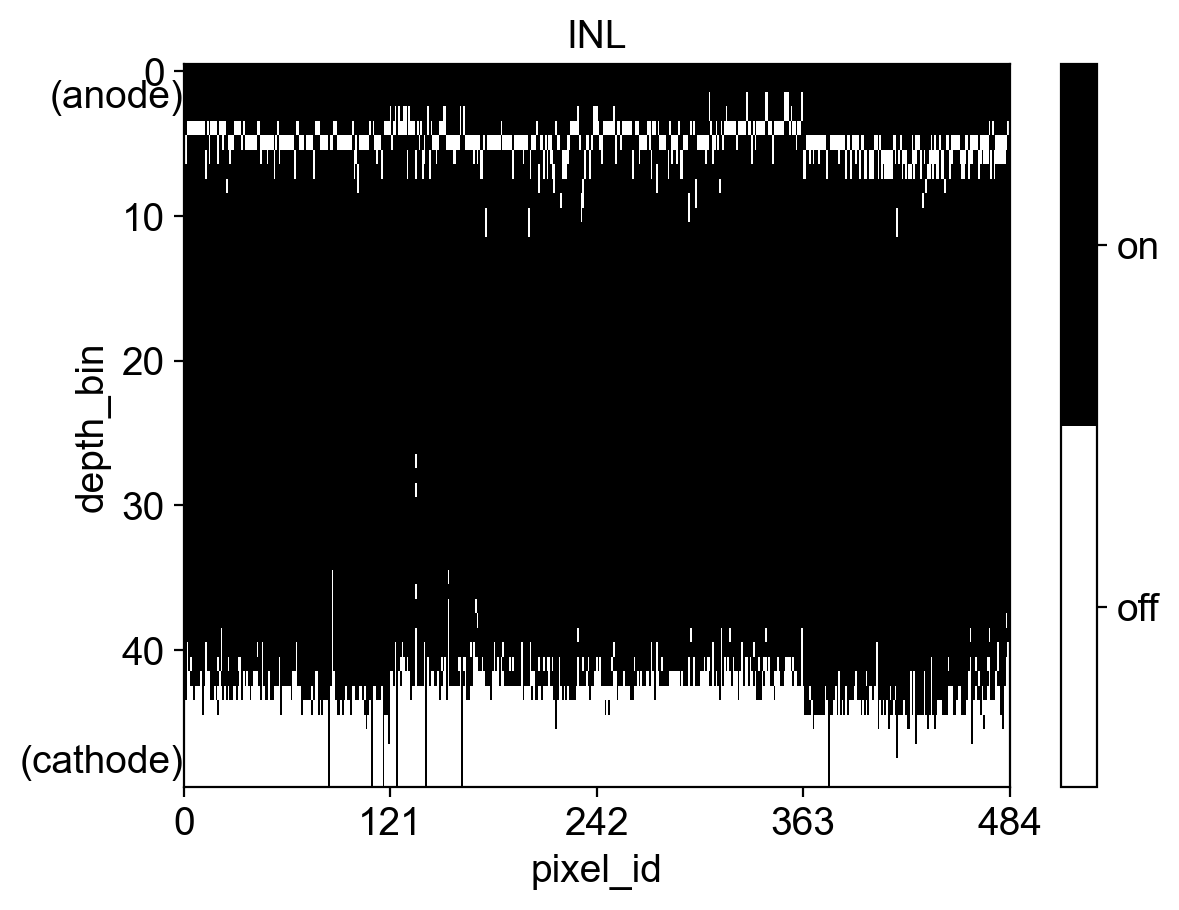}
    \includegraphics[width=0.32\linewidth]{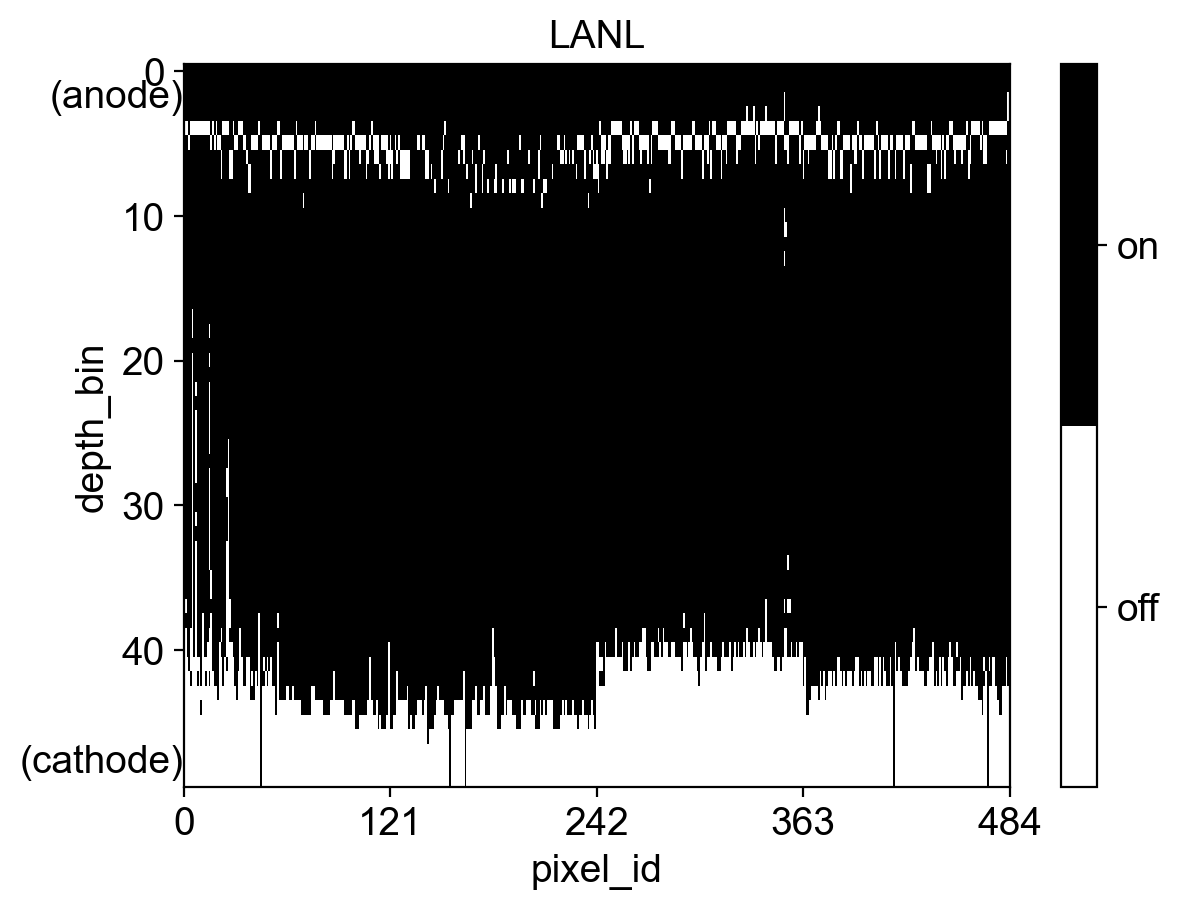}\\
    \includegraphics[width=0.32\linewidth]{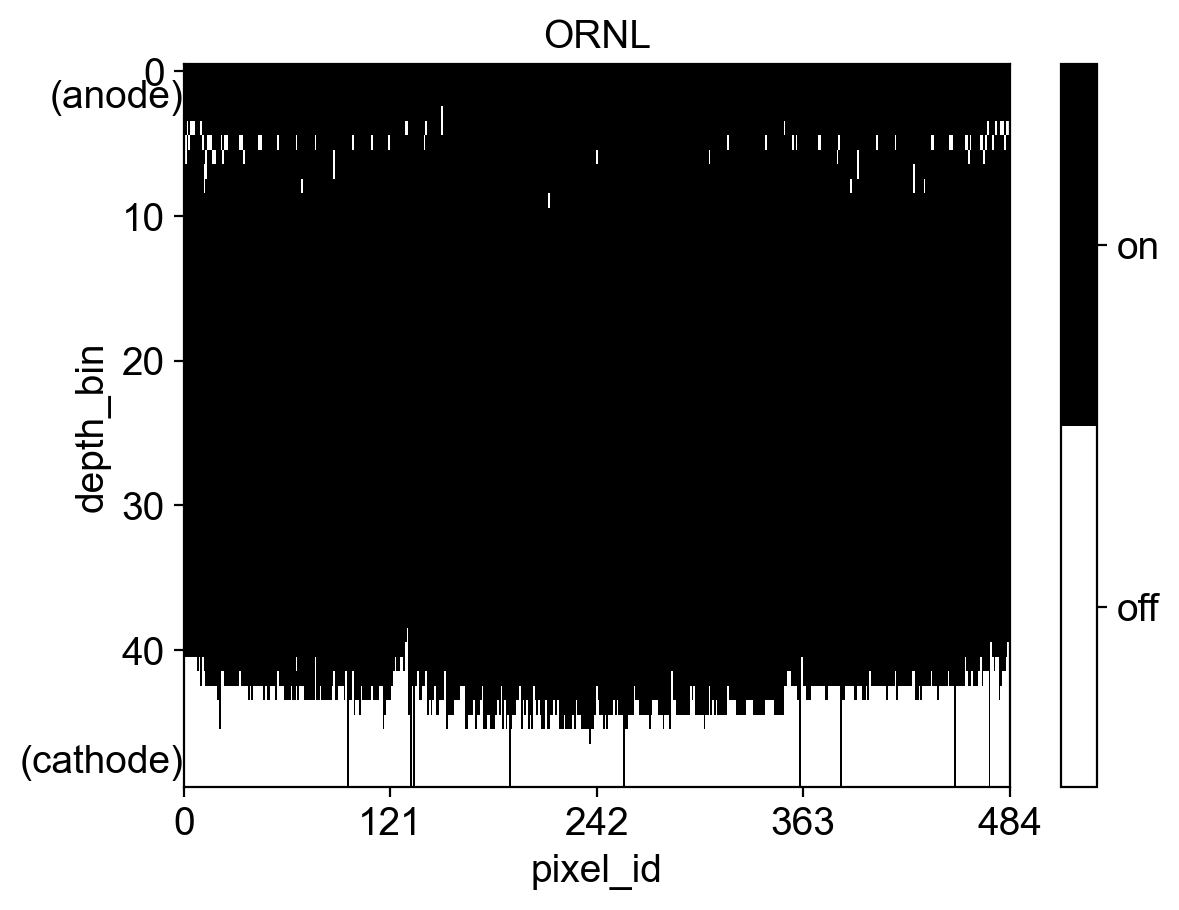}
    \includegraphics[width=0.32\linewidth]{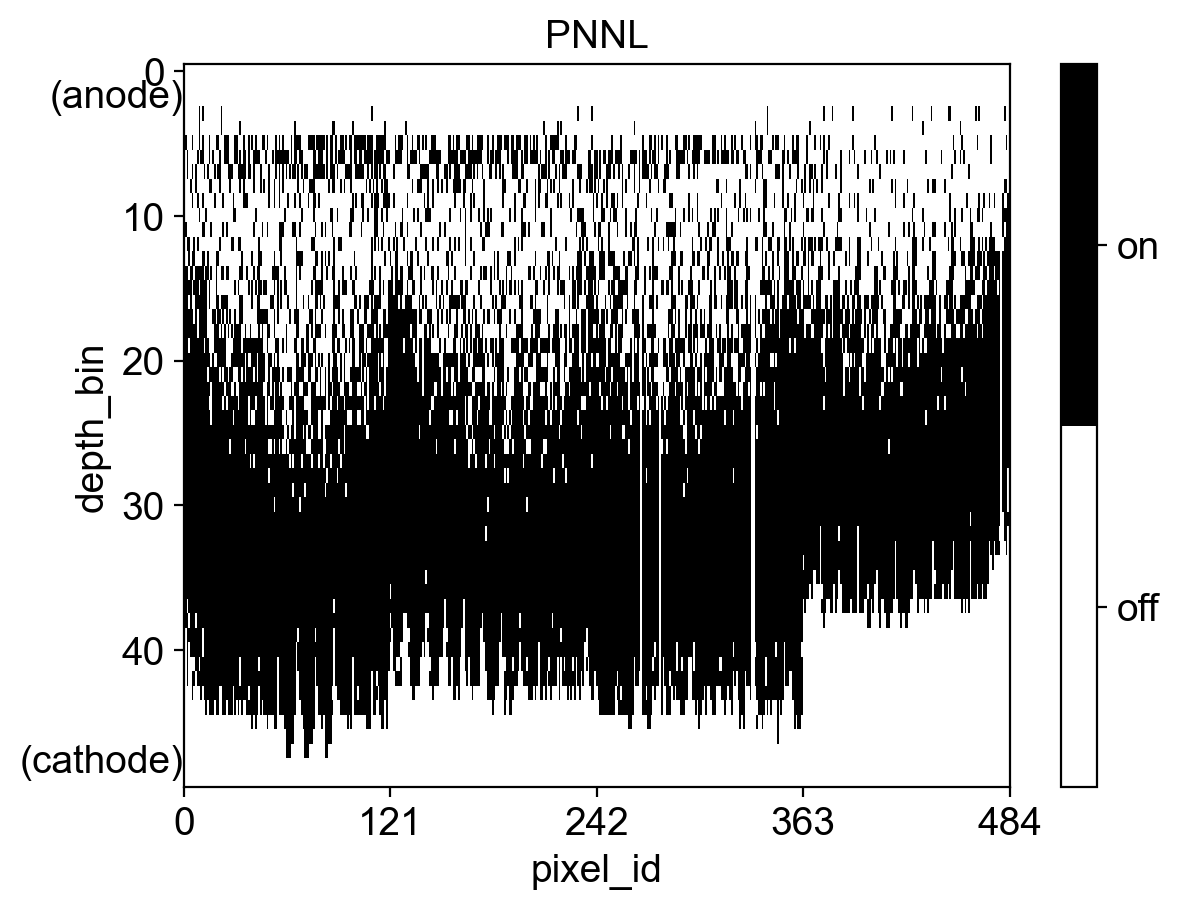}
    \includegraphics[width=0.32\linewidth]{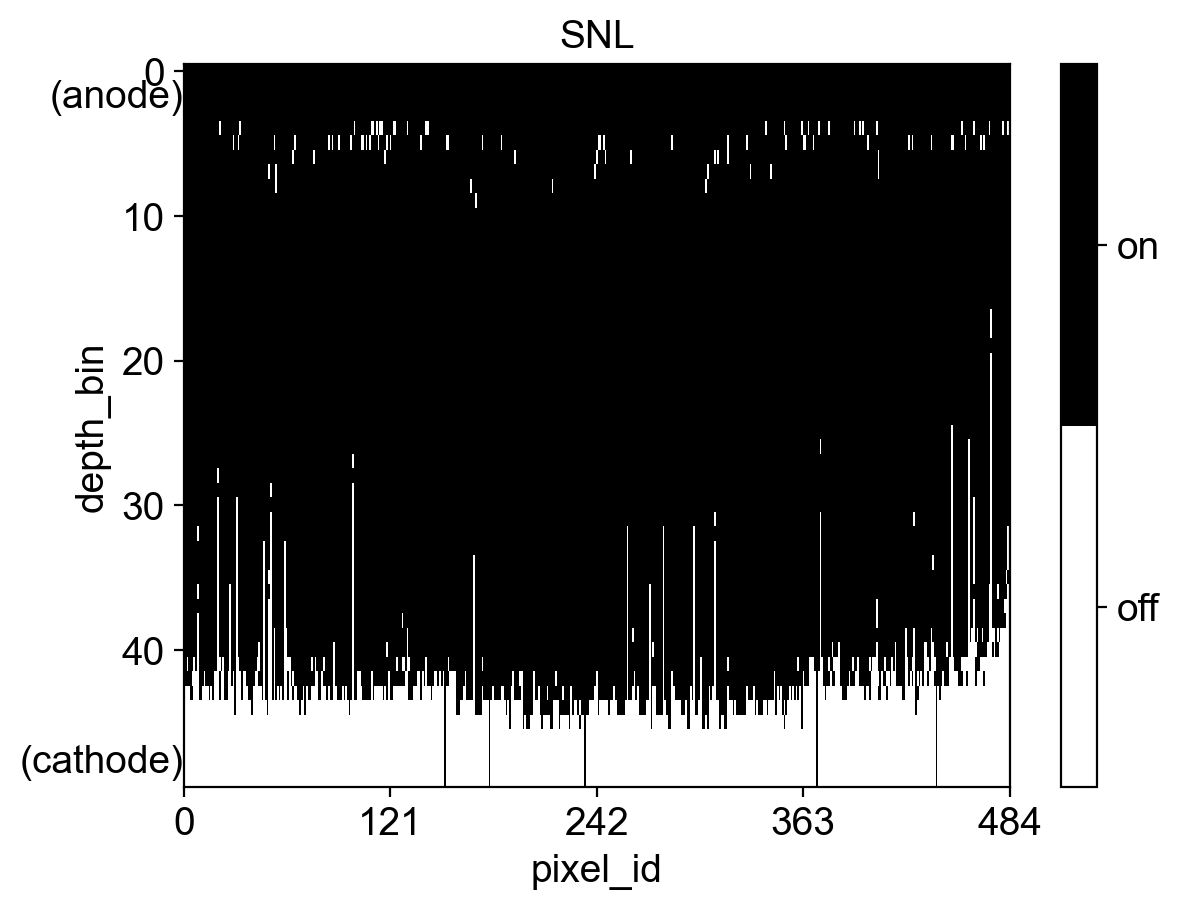}
    \caption{
        Best cluster mask from each detector optimization.
    }
    \label{fig:best_masks}
\end{figure*}

Similar or better {\tt spectre-ml} + {\tt pyGEM} improvements in the U-235 enrichment relative uncertainty are observed across all six M400 detectors tested---see Table~\ref{tab:summary}.
The metric relative improvements range from $12\%$ to $26\%$, with a mean of $20\%$.
Four out of six detectors are best improved by a GaussianMixture clusterer with $2$ out of $3$ clusters retained, while the PNNL detector removes one additional cluster and the BNL detector sees best results from the random depth bin clusterer---see the best cluster masks in Fig.~\ref{fig:best_masks}.
These six masks, though found for different detectors via different model parameters, all remove largely depth-based clusters, mostly near the cathode and/or anode.
CZT resolution degradation near the anode is well-known~\cite{li1999spatial}, and worse signal-to-background ratios are observed near the cathode in Fig.~\ref{fig:depth_spectra}.
See also the discussion on the similarity of the six masks in Section~\ref{sec:discussion}.
The relative efficiencies after voxel cluster removal range from $54\%$ (PNNL) to $70\%$ (BNL), with the remaining four detectors more tightly grouped between $61\%$ and $68\%$.
All optimizations produce very small FWHM degradations, on average changing from $1.15\%$ (bulk) to $1.21\%$ (best), for reasons that are not yet known.
Finally, the {\tt pyGEM} fit quality improves in all six optimizations, from an average {\tt Nqfit} of $2.81$ to $1.71$.

\begin{table*}[!htbp]
    \centering
    \caption{
        Summary of {\tt spectre-ml} + {\tt pyGEM} U-235 enrichment assay optimization results across six M400 detectors.
    }
    \begin{tabular}{c||c|c|c|c|c|c||c}
         detector & BNL & INL & LANL & ORNL & PNNL & SNL & mean \\\hline\hline
         best model ($\ncomp; \nclus$) & --- & $6; 2 \text{ of } 3$ & $5; 2 \text{ of } 3$ & $2; 2 \text{ of } 3$ & $4; 1 \text{ of } 3$ & $4; 2 \text{ of } 3$ & --- \\
         186 keV rel.~eff. & $70\%$ & $62\%$ & $61\%$ & $68\%$ & $54\%$ & $65\%$ & $63\%$ \\\hline
         U-235 wt\%, bulk & $1.974 \pm 0.020$ & $1.935 \pm 0.020$ & $1.951 \pm 0.020$ & $1.938 \pm 0.019$ & $1.956 \pm 0.021$ & $1.875 \pm 0.021$ & $1.938 \pm 0.008$ \\
         U-235 wt\%, best & $1.959 \pm 0.018$ & $1.937 \pm 0.015$ & $2.004 \pm 0.016$ & $1.950 \pm 0.014$ & $1.962 \pm 0.018$ & $1.901 \pm 0.016$ & $1.952 \pm 0.007$ \\\hline
         metric improvement & $13\%$ & $24\%$ & $21\%$ & $24\%$ & $12\%$ & $26\%$ & $20\%$ \\\hline
         186 keV FHWM, bulk & $1.12\%$ & $1.13\%$ & $1.15\%$ & $1.20\%$ & $1.18\%$ & $1.10\%$ & $1.15\%$ \\
         186 keV FHWM, best & $1.14\%$ & $1.21\%$ & $1.24\%$ & $1.32\%$ & $1.20\%$ & $1.14\%$ & $1.21\%$ \\\hline
         {\tt pyGEM} fit quality, bulk & $2.78$ & $2.78$ & $2.73$ & $2.47$ & $2.70$ & $3.37$ & $2.81$ \\
         {\tt pyGEM} fit quality, best & $2.05$ & $1.55$ & $1.62$ & $1.37$ & $1.93$ & $1.76$ & $1.71$
    \end{tabular}
    \label{tab:summary}
\end{table*}

\begin{figure*}[!htbp]
    \centering
    \includegraphics[width=0.49\linewidth]{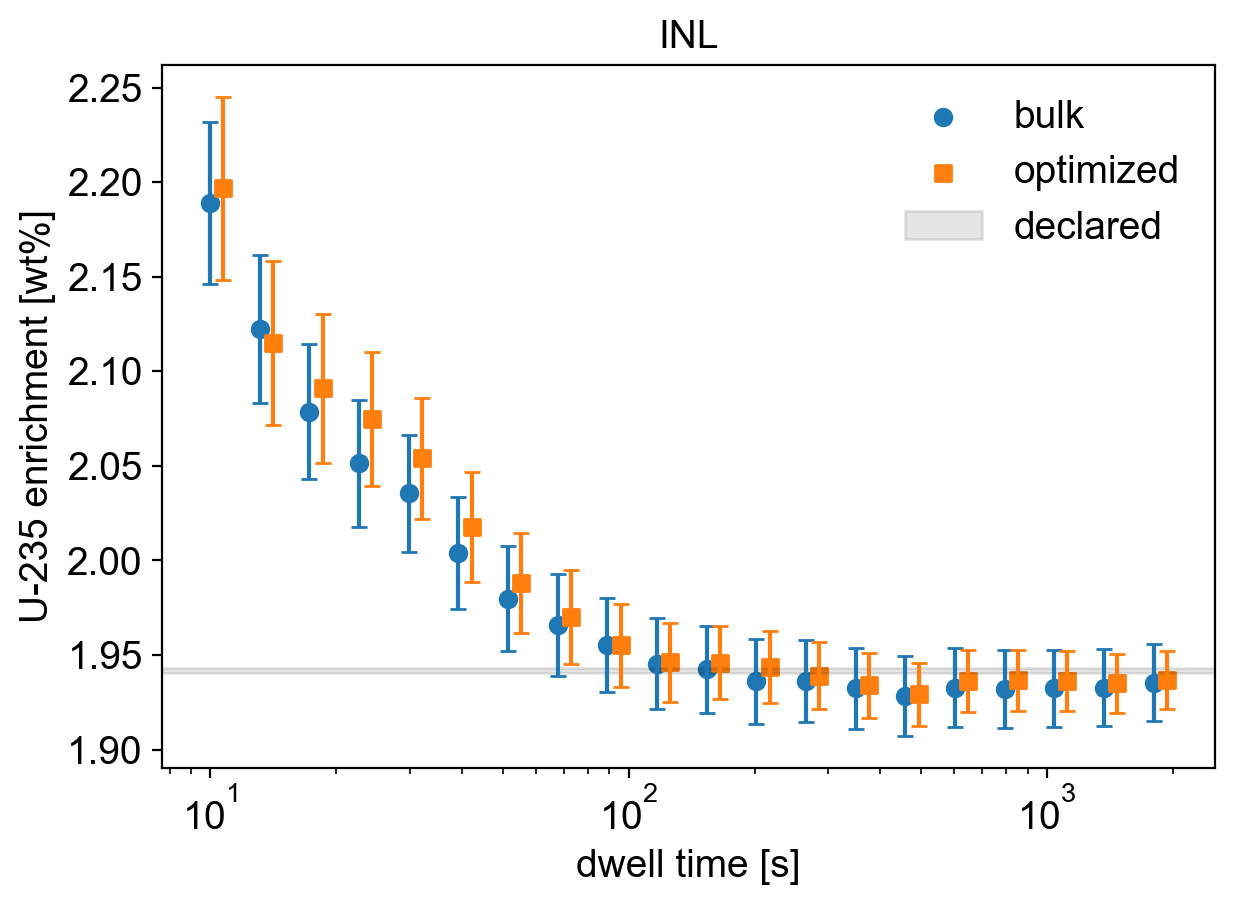}
    \includegraphics[width=0.49\linewidth]{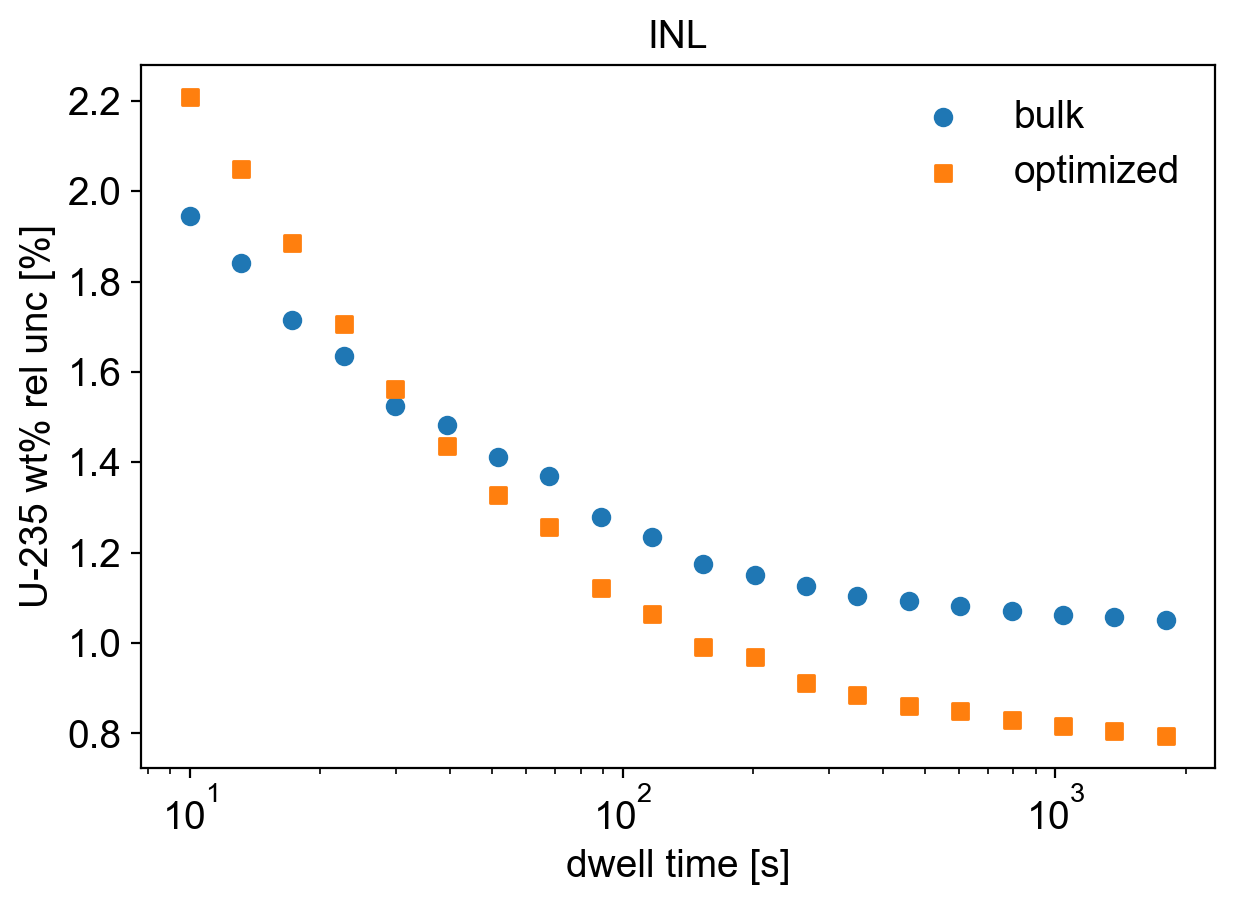}\\
    \includegraphics[width=0.49\linewidth]{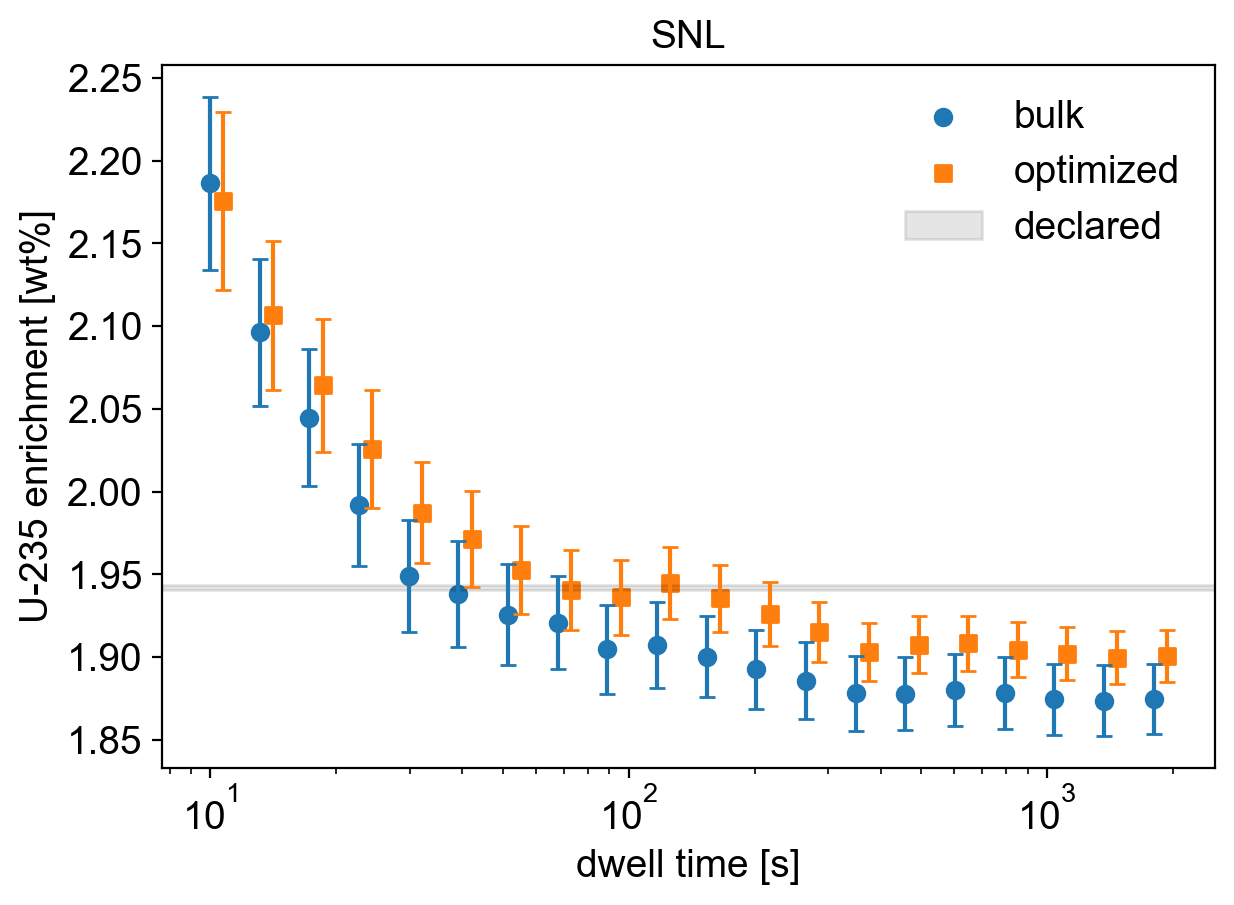}
    \includegraphics[width=0.49\linewidth]{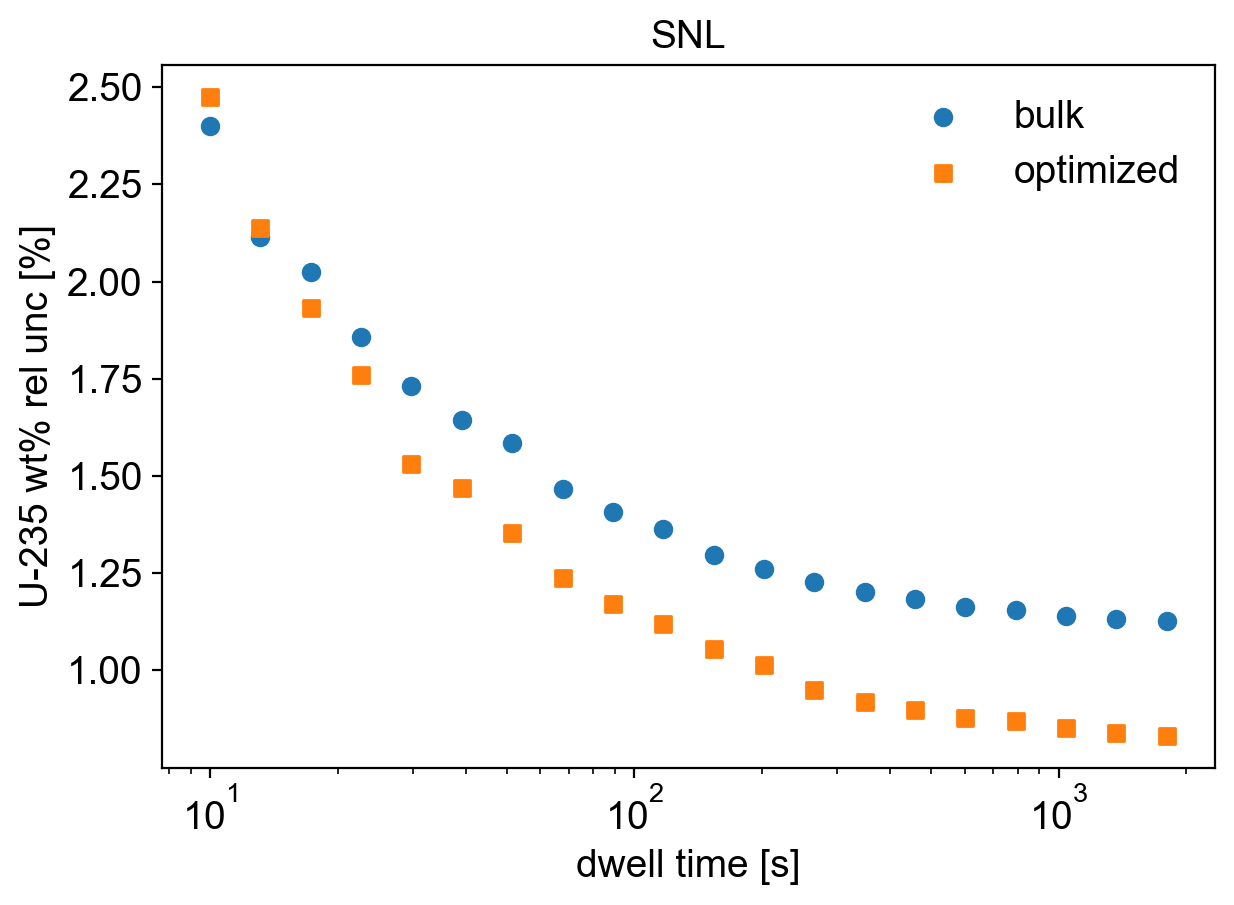}\\
    \includegraphics[width=0.49\linewidth]{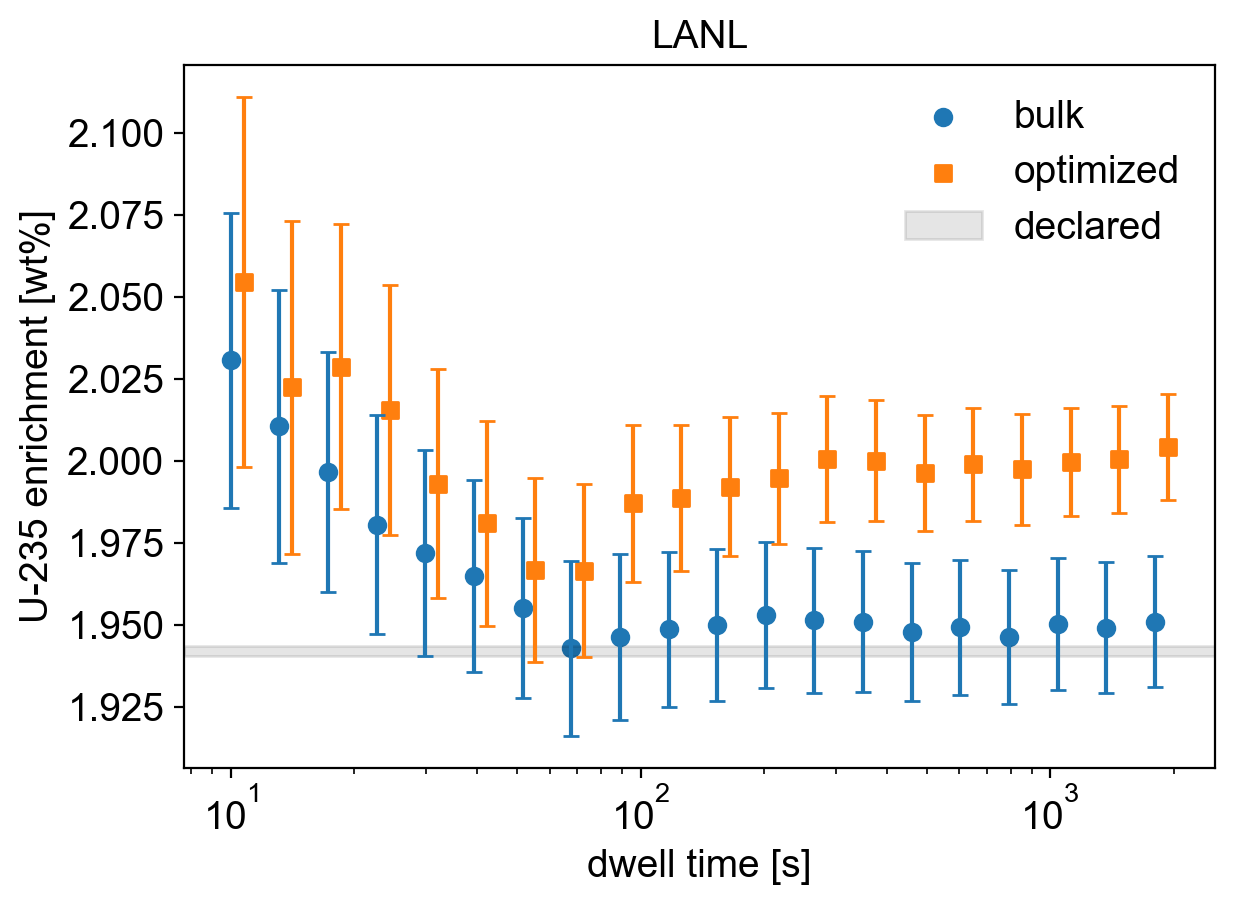}
    \includegraphics[width=0.49\linewidth]{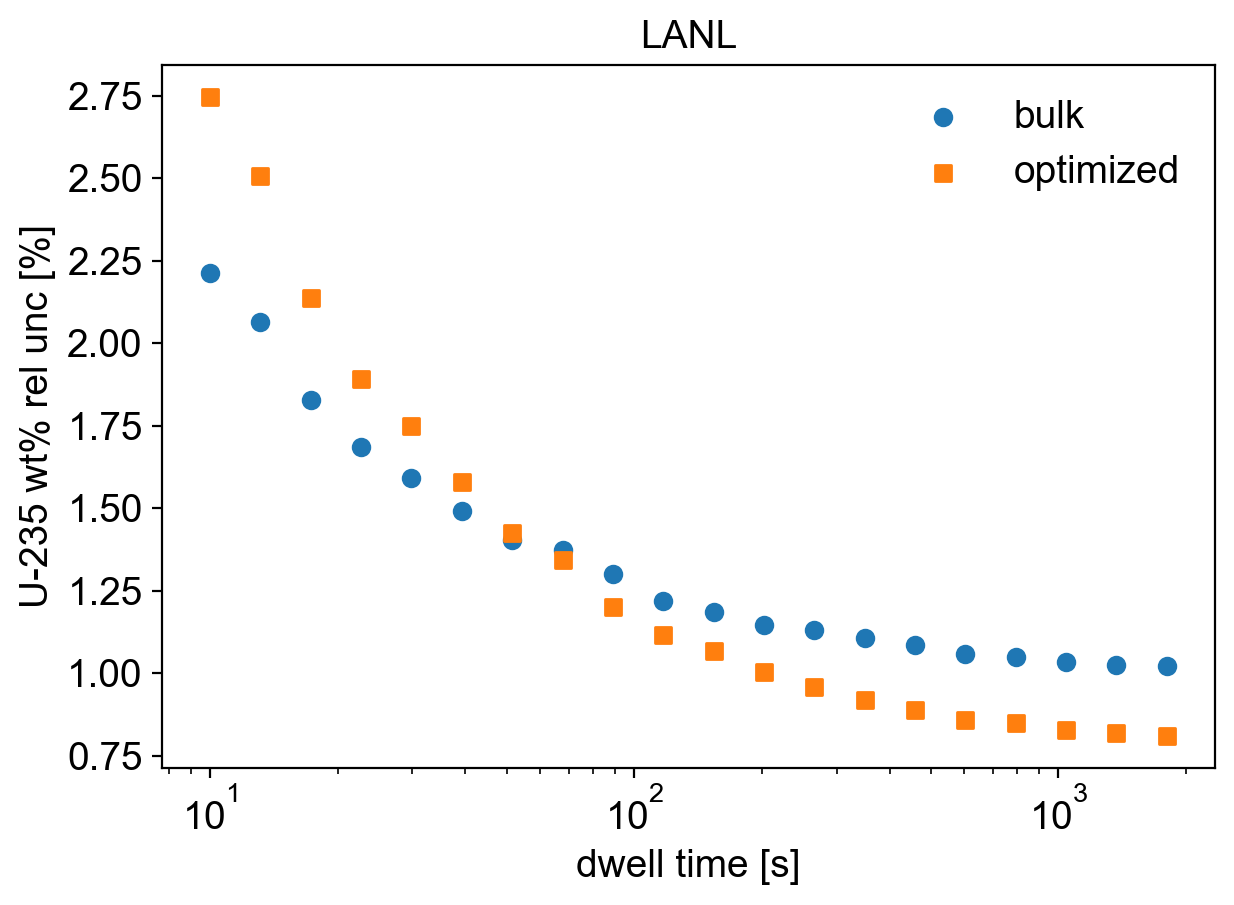}
    \caption{
        Trends in {\tt pyGEM} analysis metrics vs dwell time $t$, using the optimum long-dwell-computed mask from the INL (top row), SNL (middle row), and LANL (bottom row) detectors.
        Left column: U-235 enrichment (wt\%).
        The optimized (orange) points are slightly offset along the $x$-axis from their corresponding bulk (blue) points for visual clarity.
        The gray band shows the SRM~969 certification of $1.9420 \pm 0.0014$~wt\%~\cite{srm969}.
        Right column: U-235 relative uncertainty~(\%).
    }
    \label{fig:trends_vs_time}
\end{figure*}

The analyses above were run with the full $30$-minute dwell time datasets, where systematic fit uncertainty is expected to dominate over statistical counting uncertainty.
To expand this analysis, we also test the performance of these long-dwell-computed masks on shorter sub-samples of the same data.
Fig.~\ref{fig:trends_vs_time} shows the U-235 enrichment (wt\%) and its relative uncertainty (the direct optimization metric) as a function of dwell time $t$ for three of the six detectors.
As expected, the relative uncertainty decreases with increasing dwell time, for both the bulk and optimized results.
The \textit{initial} optimized relative uncertainty tends to be larger than the initial bulk value, presumably due to worse counting statistics from the lower efficiency, but the rate of decrease in the optimized curves is faster, leading to a \textit{crossover time}, $t^*$, where the optimized relative uncertainty first falls below the bulk result.
As shown in Table~\ref{tab:times}, these crossover times range from $17$~s to $117$~s, with an average of $49$~s, depending on the detector.
Similarly, we can quantify the \textit{speedup factor}, $f_t$, that could be achieved by dwelling with the optimized model only until the relative uncertainty from the bulk dataset at time $t$ is reached.
For the full $30$-minute datasets, these speedups $f_{30}$ range from $5.2 \times$ to $15.4 \times$, with an average of $10.3\times$, and are also given in Table~\ref{tab:times}.
For a shorter $5$-minute dwell time, the speedups $f_5$ range from $2.3\times$ to $5.2\times$, with an average of $3.8\times$, indicating that the optimized models can achieve the same relative uncertainty as a $5$-minute bulk measurement in about a quarter of the time.

\begin{table}[!htbp]
    \centering
    \caption{Summary of time efficiency improvements}
    \begin{tabular}{c||c|c|c}
        detector & crossover time $t^*$ [s] & speedup factor $f_{30}$ & $f_{5}$ \\\hline\hline
        BNL & 23 & 8.9 & 3.0 \\
        INL & 39 & 11.7 & 5.2 \\
        LANL & 68 & 8.9 & 3.0 \\
        ORNL & 30 & 11.7 & 3.9 \\
        PNNL & 117 & 5.2 & 2.3 \\
        SNL & 17 & 15.4 & 5.2 \\\hline
        mean & 49 & 10.3 & 3.8
    \end{tabular}
    \label{tab:times}
\end{table}

\begin{figure}[!htbp]
    \centering
    \includegraphics[width=0.99\linewidth]{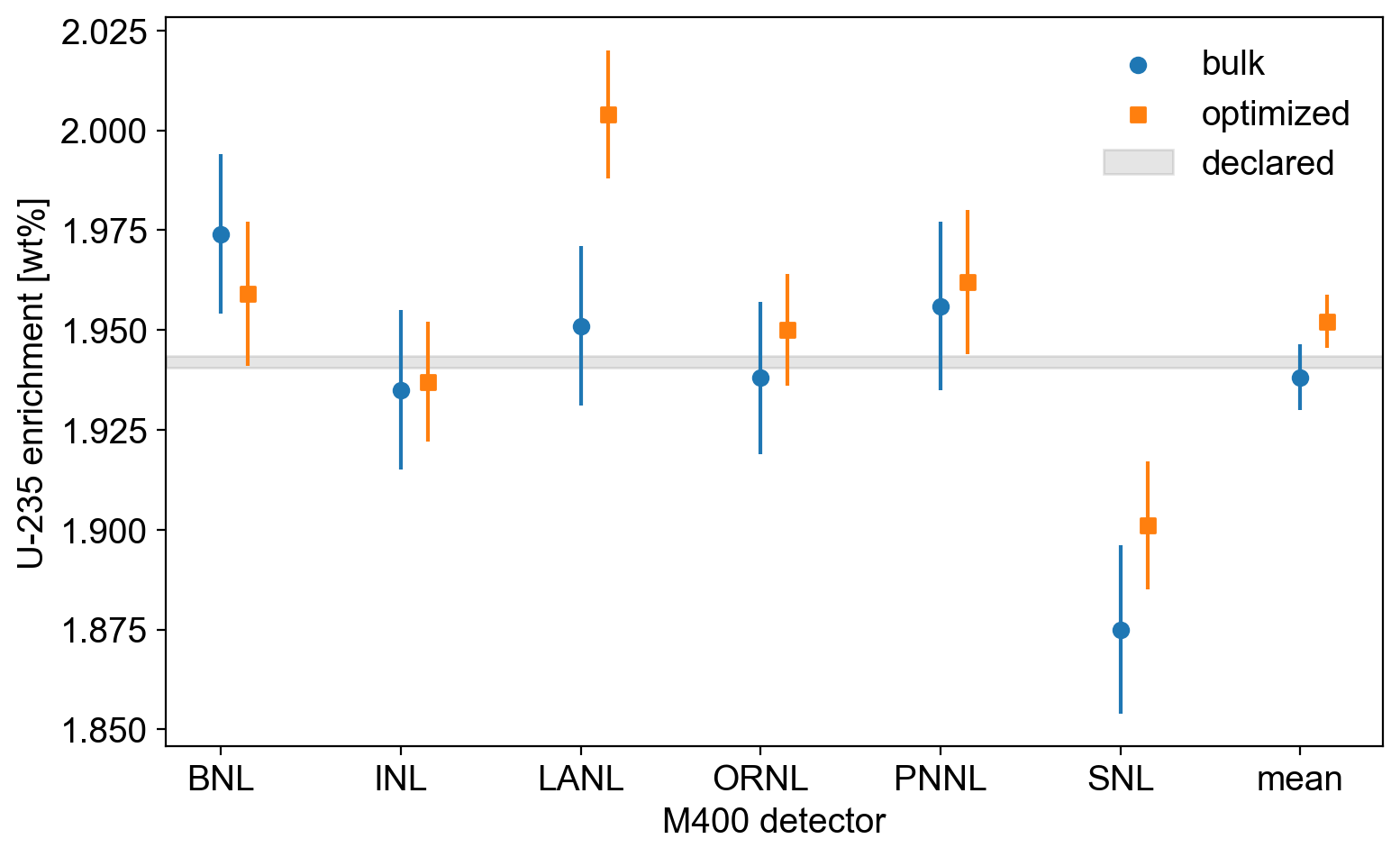}
    \caption{
        Summary of U-235 enrichment assay results with (orange squares) and without (blue circles) the {\tt spectre-ml} + {\tt pyGEM} optimization.
    }
    \label{fig:enrichment_summary}
\end{figure}

In addition to the improvement in the relative uncertainties in Fig.~\ref{fig:trends_vs_time}---the precisions of the enrichment calculations---it is also important to discuss the effects on the accuracies.
The INL, SNL, and LANL detector datasets in Fig.~\ref{fig:trends_vs_time} were specifically chosen to highlight three different cases: the bulk detector accuracy is essentially unchanged by the optimization (INL); it is slightly improved by the optimization (SNL); and it is slightly degraded by the optimization (LANL).
Similarly, Fig.~\ref{fig:enrichment_summary} summarizes the U-235 enrichment assay results for all six detectors.
The LANL M400 result appears to be an outlier that pulls the six-detector mean above the sample's declared enrichment band, but other than the LANL detector, each detector result agrees within error bars.
Three of the six detectors have computed U-235 enrichments closer to the declared value without using the {\tt spectre-ml} optimization, and three of six are closer with it.
With or without the {\tt spectre-ml} optimization, the computed enrichments can fall on either side of the declared value---but with the optimization, the uncertainties are improved without any obvious bias.

\section{Discussion}\label{sec:discussion}

The results of Section~\ref{sec:results} demonstrate that {\tt spectre-ml} can improve uranium enrichment assay uncertainties found via the {\tt pyGEM} code, thereby offering faster measurement times for in-field safeguards inspections.
Here we discuss a handful of remaining limitations, and present opportunities for future work.

First, the {\tt spectre-ml} parameter sweeps are much smaller than those used in Ref.~\cite{vavrek2025data}, using only $2$--$6$ clusters and NMF components, only the Gaussian Mixture clusterer (among the ML options), and no NMF regularization.
It is likely that the observed $20\%$ average relative improvement in the U-235 enrichment relative uncertainty metric would improve if the parameter sweeps were expanded.
Currently, we are compute-limited by the worse performance of the underlying {\tt scikit-learn} libraries on the Windows machine required for the {\tt pyGEM} code, with even these smaller parameter sweeps taking ${>}1$~hr wall time on a $4$-core 11\textsuperscript{th} Gen Intel Core i7-1185G7 processor.
Performance is also degraded by some difficult-to-avoid array copies in the {\tt pyGEM} pipeline, as well as the need to run both NMF and fitting routines on spectra of $O(1000)$ bins for the entire U-235 region rather than the $O(100)$ bins for a single photopeak.
Better parallelization, either in the {\tt scikit-learn} routines or over individual parameter combinations, could greatly improve the speed of the optimization.

We also note that the observed $20\%$ average relative improvement is smaller than the ${\sim}3\times$ improvement observed in the U-235 peak analysis of Ref.~\cite{vavrek2025data}, as expected, due to the much lower systematic fit error arising from the more complex spectrum fit (which includes the $195$~keV peak) in {\tt pyGEM}.
This re-emphasizes the importance of carefully defining the performance metric to be used to avoid specification gaming.

The similarity of four out of six best voxel removal models in Table~\ref{tab:summary} and therefore masks in Fig.~\ref{fig:best_masks} suggests that a single common mask could be applied to improve many M400 detectors without having to specifically optimize each detector, even despite inter-detector performance differences.
It would be valuable in the future to quantify this level of mask transferability among detectors.

Our analysis here has been limited to using the $4.46\%$-enriched U-235 standard as a calibration standard for assaying the $1.94\%$-enriched sample.
In the future, it could be interesting to repeat the analysis for the reverse combination, and to include the $20.11\%$-enriched measurements from the same collection of datasets.
Similarly, other safeguards measurement scenarios such as UF$_6$ cylinders could be considered.

Finally, we note that the ease with which {\tt pyGEM} was integrated with {\tt spectre-ml} bodes extremely well for future integrations of other analysis codes and/or other detector arrays.
For instance, {\tt FRAM}~\cite{sampson1989fram} could be integrated to test the ability of {\tt spectre-ml} to improve the challenging analysis of multiple densely-packed photopeaks in Pu spectra.

\section{Conclusions}
We have integrated the {\tt pyGEM} uranium enrichment analysis code with the {\tt spectre-ml} spectroscopic optimization framework, and shown that the uranium enrichment relative uncertainty can be directly used as an optimization target.
This {\tt spectre-ml} + {\tt GEM} workflow provides a $20\%$ relative improvement (averaged over six H3D M400 detectors) in the U-235 enrichment relative uncertainty in $30$-minute long measurements, and retains similar enrichment accuracy to the unoptimized {\tt pyGEM}-only results.
These reductions in enrichment relative uncertainty can lead to significant time savings for in-field nuclear safeguards inspections, with the optimized results achieving the same overall relative uncertainty as the unoptimized $30$-minute measurements around $10\times$ faster.
Future work could integrate other safeguards NDA software such as {\tt FRAM} with {\tt spectre-ml} for improved Pu spectroscopy, and investigate whether a single common optimization result can be applied across multiple detectors to provide improved spectroscopy with minimal operational cost.

\appendix\label{sec:code_appendix}
The following pseudocode provides a basic outline of the application programming interface (API) that an end-user would interact with to integrate their own analysis code with the {\tt spectre-ml} framework.

\begin{lstlisting}
# spectre-ml framework: abstract classes
class RankingMetric(ABC):
    ...
class SpectrumAnalyzer(ABC):
    ...

# user code: concrete becquerel workflow
class BecquerelSpectrumAnalyzer(SpectrumAnalyzer):
    def analyze(self):
        compute_single_peak_fits()
class BecquerelRankingMetric(RankingMetric):
    def calc(self) -> float:
        return fit_param_rel_unc

# user code: two concrete pyGEM workflows
class GEMSpectrumAnalyzer(SpectrumAnalyzer):
    def analyze(self):
        compute_activity_calibration()
        compute_U235_region_fits()
class GEMEnrichmentMetric(RankingMetric):
    def calc(self) -> float:
        return enrichment_rel_unc
class GEMFittingMetric(RankingMetric):
    def calc(self) -> float:
        return fit_chi_squared

# user code: example GEM analysis
ms = spectre_ml.ModelSelector(
    spectrum_analyzer=GEMSpectrumAnalyzer(),
    ranking_metric=GEMEnrichmentMetric(),
    voxel_spectra=voxel_spectra,
    n_clusters=[2, 3],
    n_nmf_components=[2, 3],
    ...
)
ms.fit_models()
results = ms.evaluate_models()
results.plot()
    
\end{lstlisting}

\section*{Acknowledgments}
The authors thank Michael Streicher and David Goodman (H3D, Inc.) for useful discussions.

The U.S.\ Government retains, and the publisher, by accepting the article for publication, acknowledges, that the U.S.\ Government retains a non-exclusive, paid-up, irrevocable, world-wide license to publish or reproduce the published form of this manuscript, or allow others to do so, for U.S.\ Government purposes.

\bibliographystyle{IEEEtran}
\bibliography{biblio}

\end{document}